\documentclass[a4paper,11pt]{article}
\usepackage{amsfonts,amsmath,amssymb,bm,url}
\usepackage{graphicx}
\usepackage{epstopdf}
\usepackage{xcolor}  
\usepackage{float}
\usepackage{graphicx}
\usepackage[utf8]{inputenc}
\usepackage{epsfig}
\usepackage{amssymb}

\title{Constraining dense matter physics using\\
 $f$-mode oscillations in neutron stars}
 
\author{Sukrit Jaiswal  \\
Indian Institute of Science Education and Research, \\
Dr. Homi Bhabha Road, Pashan, \\
Pune 411008, India \\
	\and 
	Debarati Chatterjee 
	\thanks{debarati@iucaa.in} $\>$ 	\\
    Inter-University Centre for Astronomy and Astrophysics, \\
Pune University Campus,\\
Pune - 411007, India
	}
\date{\today}

\begin{document}

\maketitle

\begin{abstract}
In this undergraduate project, $f$-mode oscillations in neutron stars are used to constrain the equation of state of dense matter. For the first time, a systematic investigation of the role of nuclear saturation parameters on the mode oscillations is performed. It is found that the uncertainty in the determination of effective nucleon mass plays the most significant role in controlling the $f$-mode frequencies. Correlations of the frequencies with astrophysical observables relevant for asteroseismology are also investigated. Future detection of $f$-mode frequencies could then provide a unique way of constraining nuclear empirical parameters and therefore the behaviour of dense matter.
\end{abstract}

\section{Introduction}
\label{sec:intro}
With the direct detection of gravitational waves (GW) from binary merger (GW170817) of neutron stars (NSs) \cite{Abbott2017a}, a new window of opportunity has opened up to directly probe their interior composition. In conjunction with astrophysical observations using multi-wavelength space-based and ground-based telescopes, gravitational wave detectors now introduce the possibility of multi-messenger astronomy, from which several astrophysical NS observables can be derived \cite{Abbott2017b}.
\\

Neutron stars are compact remnants of the evolution of massive stars, associated with core collapse supernovae. These objects have extreme densities up to several times normal nuclear matter density in the core, orders of magnitude beyond matter densities possible to investigate with terrestrial experiments. It is therefore inevitable to resort to theoretical models, in order to extrapolate the known physics at nuclear saturation to higher densities and isospin asymmetries. This extrapolation leads to large uncertainties, reflected by the proliferation of models. Different schemes (ab-initio, phenomenological) have been applied to describe such equations of state (EoS) \cite{OertelRMP}. The ab initio many-body models use realistic nucleon-nucleon (NN) interaction which is obtained by fitting the NN scattering data. The phenomenological models are based on nuclear density functional theories with effective nucleon-nucleon interactions, which are calibrated by reproducing the nuclear bulk properties around saturation density. Different models show different high density behaviour, resulting in different predictions for NS observables, thus making it difficult to interpret and compare directly with astrophysical data.
\\

NSs are visible throughout the electromagnetic spectrum. A number of global properties of NSs, such as its mass and radius, can be deduced from multi-wavelength astronomical data.  Accurate estimations of NS masses using Post-Newtonian effects in relativistic NS binaries indicate maximum masses to be close to 2~$M_{sol}$ \cite{Demorest,Antoniadis}. Radius estimates obtained from X-ray data are less precise, but the recently launched NICER mission \cite{NICER} is soon expected to measure radii with a precision of up to 5\%. These observables can be obtained from the NS EoS by solving equations for hydrostatic equilibrium. Comparison with astrophysical data then allows us to put important constraints on the EoS models and consequently on the nature of dense matter.
\\

Gravitational waves (GW) are considered one of the most promising tools for constraining dense matter physics, as they can directly probe the interior composition of neutron stars. Quasi-normal modes may be excited in oscillating NSs, producing copious amounts of GWs. These modes, such as fundamental modes ($f$-modes), pressure modes ($p$-modes), buoyancy $g$-modes, rotational $r$-modes, are classified according to the restoring forces that bring the system back to equilibrium. The most exciting fact is that the mode frequencies and damping timescales contain signatures of the interior composition of neutron stars. 
\\

GWs may be emitted from NSs, both isolated or in binary. It was shown that during a merger, the NSs in binary exert strong tidal forces on one another, and the resulting deformation depends on their compactness \cite{Abbott2017a}. This can lead to important constraints on stellar radii and consequently on the dense matter EoS \cite{Abbott2018}. Further, quasi-normal modes may also be excited during the merger and post-merger phases \cite{Abbott_p-g}. Particularly interesting are the $f$-modes, as they are expected to have frequencies of $\sim$ 1-2 kHz and produce generous amounts of gravitational radiation through the CFS (Chandrasekhar-Friedman-Schutz) mechanism when unstable \cite{Glampedakis}. 
\\

It is therefore of great interest to constrain the NS EoS using studies of $f$-modes. The goal of NS asteroseismology is to express the $f$-mode frequency or damping timescale in terms of global NS observables, independent of the underlying EoS. Detection of $f$-modes would then allow us to invert such relations and obtain constraints for the NS EoS \cite{Andersson96,Andersson98,Schutz}.  However, most of such studies are confined to polytropic or parametrized EoSs due to their simplicity \cite{Andersson96,BaoAnLi,Salcedo}. There are a few investigations that considered realistic equations of state, but with only a few representative parameter sets \cite{Andersson98,Schutz,Doneva2013}. Although the dispersion of frequencies due to different EoSs is evident, it is impossible to compare between the chosen EoSs as they correspond to very different nuclear matter properties. Further, the EoSs chosen do not span the entire parameter space of empirical quantities allowed by state-of-the-art nuclear experimental data. Fits using such EoSs have been used to derive relations between mode frequencies and global variables (mean density, compactness). It is thus very difficult to extract any direct information about dense matter physics from such correlations. 
\\

In this work, which was done as a short exploratory undergraduate project, we investigate the influence of the underlying nuclear matter properties on $f$-mode frequencies. Within the framework of the Relativistic Mean Field (RMF) model, we systematically explore the parameter space allowed by current nuclear experimental data and perform studies of the sensitivity of the mode frequencies to the uncertainties in each of these parameters. We also investigate the role of these uncertainties on the fit relations relevant for asteroseismology of non-rotating NSs.
\\

In Sec.~\ref{sec:formalism}, we describe the microscopic calculation of the EoS as well as the determination of macroscopic NS observables from such an EoS. We also elaborate on the pulsation equations to be solved to obtain $f$-mode frequencies in non-rotating NSs. We present the results of our investigations in Sec.~\ref{sec:res}, including the sensitivity study and asteroseismology relations. We summarize our findings in Sec.~\ref{sec:disc}. 
\\

\section{Formalism}
\label{sec:formalism}

\subsection{Microscopic description}
\label{sec:micro}
 As already mentioned in Sec.~\ref{sec:intro}, we perform this investigation within the framework of the RMF model \cite{HorowitzSerot}. Such models have already been applied successfully to describe nuclear matter and nuclei \cite{OertelRMP}. In order to obtain the EoS of nuclear matter in the RMF Model, we start from the following interaction Lagrangian density:
\begin{eqnarray}
{\cal{L}}_{int} &=& \sum_N \bar{\Psi_N} \left[ g_{\sigma} \sigma - g_{\omega} \gamma^{\mu} \omega_{\mu} - \frac{g_{\rho}}{2} \gamma^{\mu} \vec{\tau} \vec{\rho_{\mu}} \right] \Psi_N \nonumber \\
&-& \frac{1}{3} b m (g_{\sigma} \sigma)^3 -\frac{1}{4} c  (g_{\sigma} \sigma)^4 \nonumber \\
&+& \Lambda_{\omega} (g_{\rho}^2 \vec{\rho_{\mu}} \vec{\rho^{\mu}} ) (g_{\omega}^2 \omega_{\mu} \omega^{\mu} ) + \frac{\zeta}{4!} (g_{\omega}^2 \omega_{\mu} \omega^{\mu} )^2
\label{eq:lagr}
\end{eqnarray}

where $\Psi_N$ is the Dirac spinor for nucleons $N$, $m$ is the vacuum nucleon mass. The interaction among the nucleons is mediated by the exchange of the scalar ($\sigma$), vector ($\omega$) and the isovector ($\rho$) mesons. The isoscalar nucleon-nucleon couplings $g_{\sigma}$ and $g_{\omega}$ are determined by fixing them to nuclear saturation properties. The $\sigma$ meson self-interaction terms $b$ and $c$ ensure the correct description of nuclear matter at saturation density. The isovector and mixed $\omega$-$\rho$ couplings $g_{\rho}$ and $\Lambda_{\omega}$ can be related to empirical quantities such as symmetry energy ($J_{sym}$) and its slope ($L_{sym}$) \cite{Horowitz2001,Fattoyev2010,Chen2014,Hornick}. The quartic $\omega$ self-coupling $\zeta$ is set to zero. We do not consider terms $\cal{O}$(3) and above in the expansion with density and asymmetry, as nuclear experimental data to constrain such parameters have large uncertainties.
\\

The uncertainty in the nuclear empirical quantities derived from saturation data is also reflected in the uncertainty in the determination of the RMF model parameters. In order to test our results, we first consider using two commonly used parametrizations \cite{GM}: GM1 and GM3, for which the EoSs are well known. Once our numerical scheme is verified, we want to explore the entire RMF parameter space defined by present uncertainties of theoretical and experimental nuclear saturation data. The range of values of empirical RMF parameters explored in this work is summarized in Table~\ref{tab:emppara}. For each individual parameter ``variation" within the ranges shown in the table, the others are kept at the ``fixed" values.
 \\
 \begin{table*}[ht]
 \centering
   \caption{Empirical parameter values for RMF models considered in this work.}
\begin{tabular}{|c|c|c|c|c|c|c|}
\hline
   Model & $n_{sat}$ & $E_{sat}$ & $K_{sat}$ & $J_{sym}$ & $L_{sym}$ & $m^*/m$ \\
    {} & ($fm^{-3}$) & (MeV) & (MeV) & (MeV) & (MeV) & {}\\
 \hline
{GM1} & 0.153 & -16.3 & 300 & 32.5 & 93.7 & 0.70 \\
\hline
{GM3} & 0.153 & -16.3 & 240 & 32.5 & 89.7 & 0.78 \\
\hline 
\hline
{RMF fixed} & {0.16} & {-16.0}  & {240} & {32} & {60} & {0.60} \\
\hline
{  variation} & {[0.15,0.16]} & {[-16.5,-15.5]}  & {[240,280]} & {[30,32]} & {[50,60]} & {[0.55,0.75]} \\
\hline
\end{tabular}
\label{tab:emppara}
\end{table*}

Given the Lagrangian density Eq.~(\ref{eq:lagr}), one can solve the equations of motion of the constituent particles as well as those of the mesons. In the mean-field approach, the meson fields are replaced by their mean-field expectation values. One then calculate the EoS (pressure - energy density relationship) using this RMF model. The energy density is given by \cite{Hornick}
\begin{eqnarray}
\varepsilon &=& \sum_N \frac{1}{8 \pi^2} \left[ k_{F_N} E_{F_N}^3 +  k_{F_N}^3 E_{F_N} - {m^*}^4 \ln \frac{k_{F_N}+E_{F_N}}{m^*} \right]  \nonumber \\
&+& \frac{1}{2} m_{\sigma}^2 \bar{\sigma}^2 + \frac{1}{2} m_{\omega}^2 \bar{\omega}^2 + \frac{1}{2} m_{\rho}^2 \bar{\rho}^2 \nonumber \\
&+& \frac{1}{3} b m (g_{\sigma} \sigma)^3 -\frac{1}{4} c  (g_{\sigma} \sigma)^4 \nonumber \\
&+& 3 \Lambda_{\omega} (g_{\rho} g_{\omega} \bar{\rho} \bar{\omega})^2 + \frac{\zeta}{8} (g_{\omega} \bar{\omega})^4~.
\end{eqnarray}
The pressure $P$ can be derived from the energy density using the Gibbs-Duhem relation \cite{GlendenningBook}
\begin{equation}
P = \sum_N \mu_N n_N - \varepsilon~,
\end{equation}
where the nucleon chemical potentials are given by
\begin{equation}
\mu_N = E_{F_N} + g_{\omega} \bar{\omega} 
+ \frac{g_{\rho}}{2} \tau_{3N} \bar{\rho}~.
\end{equation}

\begin{figure} [htbp]
\centering
\includegraphics[height=10 cm,width=10 cm]{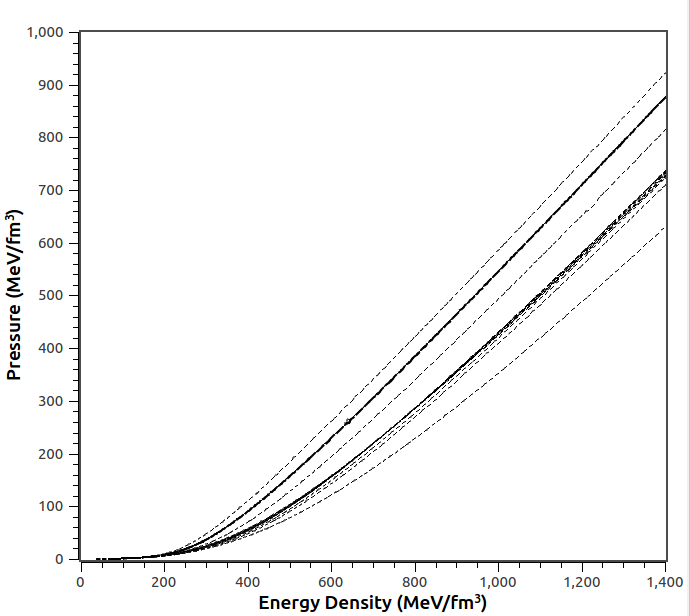}
\caption{EoSs used in this work}
\label{fig:eos_all}
\end{figure}

\subsection{Macroscopic description}
\label{sec:macro}
Given an EoS, one can obtain the macroscopic NS structure by solving the Tolman Oppenheimer Volkov (TOV) equations of hydrostatic equilibrium \cite{GlendenningBook}
\begin{eqnarray}
\frac{dm(r)}{dr} &=& 4 \pi \varepsilon(r) r^2 ~, \nonumber \\
\frac{dp(r)}{dr} &=& - \frac{[p(r) + \varepsilon(r)] [m(r)+4 \pi r^3 p(r)] }{r(r-2 m(r))}~,
\label{eq:tov}
\end{eqnarray}

Integrating the TOV equations from the centre of the star to the surface, one can obtain global NS observables, such as mass ($M$), radius ($R$) and compactness ($C=M/R$). The tidal deformability ($\Lambda$) can be obtained by solving a set of differential equations coupled with the TOV equations \cite{YagiYunes2013b}. These can then be compared to the state-of-the-art limits derived from astrophysical data, in order to impose constraints on the dense matter EoS. 
\\

\begin{figure} [htbp]
\centering
\includegraphics[height=10 cm,width=10 cm]{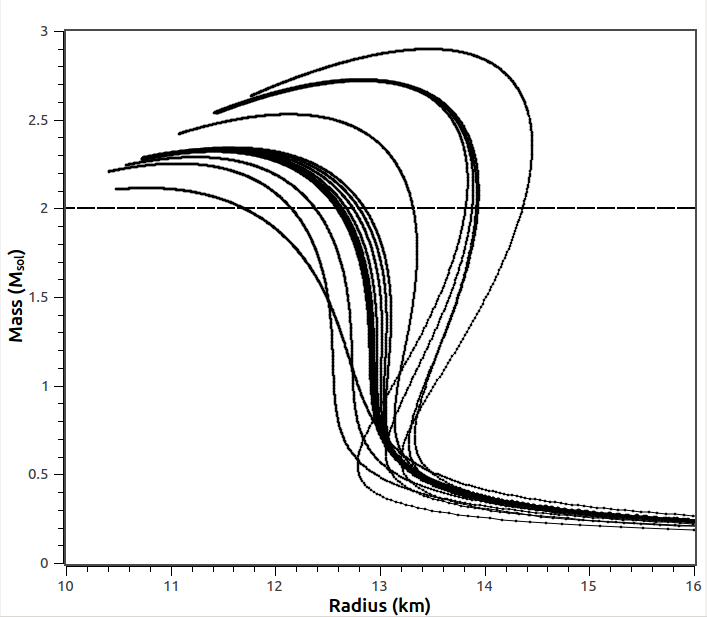}
\caption{Mass-Radius relations corresponding to the EoSs plotted in Fig.~\ref{fig:eos_all}}
\label{fig:mr_all}
\end{figure}

In Fig.~\ref{fig:eos_all}, we display all the EoS curves (pressure vs energy density in $MeV/fm^3$) considered in this work, corresponding to the parameter sets in Table~\ref{tab:emppara}. The corresponding mass-radius relations are given in Fig.~\ref{fig:mr_all}. It is evident from Fig.~\ref{fig:mr_all} that for all the EoSs considered in this work, the maximum masses lie above 2$M_{sol}$ (horizontal dotted line in the figure). 
\\


\subsection{Solving the mode pulsation equations}
The aim of this work is to investigate the influence of the uncertainty in the empirical nuclear parameters on NS $f$-modes. In general, one must solve coupled fluid and space-time perturbation equations to obtain the mode frequencies \cite{Sotani}. However, the situation simplifies if we consider weak gravitational fields and neglect the metric perturbations. This approach, known as the Cowling approximation, has been widely applied for studying Newtonian as well as relativistic NSs. Although ideally one must employ fully linearized equations in general relativity, it has been shown that the difference on applying the Cowling approximation is less than 20\% for $f$-modes \cite{Vasquez}. For this study, we only consider non-rotating NSs (See \cite{Doneva2013} for investigations of $f$-modes in rotating NSs).  
\\

Within the Cowling approximation (neglecting metric perturbations), the spherically symmetric background spacetime characterized by a line element is given by:

\begin{equation}
ds^2=-e^{2\Phi(r)} dt^2 + e^{2\Lambda(r)} dr^2 + r^2 d\theta^2 + r^2 sin(\theta)d\phi^2
\label{eq:metric}
\end{equation}

The fluid Lagrangian displacement vector is defined by
\begin{equation}
\zeta^i = (e^{-\Lambda} W, - V \partial_{\theta}, -V \sin^{-2} \partial_{\phi} )
\end{equation}
where $W(t,r)$ and $V(t,r)$ are functions of $r$ and $t$. The fluid perturbations
are decomposed into spherical harmonics $Y_{lm} (\theta, \phi)$. Using these variables, the perturbation equations for the fluid oscillations can be obtained from the conservation of the energy-momentum tensor $\delta(\nabla_{\nu} T^{\mu \nu}) = 0$. For a harmonic time dependence, the perturbation functions can be written as  $W(t,r)=W(r)e^{i \omega t}$ and  $V(t,r)=V(r)e^{i \omega t}$.
On simplification, the pulsation equations required to be solved in order to obtain these frequencies are given by \cite{Sotani}:
\begin{eqnarray}
\frac{dW(r)}{dr} &=&\frac{d\epsilon}{dP} \left[\omega^2r^2e^{\Lambda(r)-2\Phi(r)}V(r)+\frac{d\Phi(r)}{dr}W(r) \right] - l(l+1)e^{\Lambda(r)}V(r)  \nonumber \\
\frac{dV(r)}{dr} &=& 2\frac{d\Phi(r)}{dr}V(r)-\frac{1}{r^2}e^{\Lambda(r)}W(r)
\label{eq:pulseq}
\end{eqnarray} 

The functions $V(r)$ and $W(r)$, along with the frequency $\omega$, characterize the perturbation vector. We solve coupled Eq.~\ref{eq:pulseq} on a fixed background metric from the origin ($r = 0$), where the solutions behave approximately like \cite{Vasquez,Sandoval}

$$V(r)=\frac{C}{l}r^l$$
$$W(r)=Cr^{l+1}$$

where $C$ is an arbitrary constant. The other boundary condition that needs to be fulfilled is that the perturbation to the pressure must vanish at the star’s surface ($r = R$). Such condition reads

\begin{equation}
\left. \omega^2e^{\Lambda(R)-2\Phi(R)}V(R)+ \frac{1}{R^2} \left(\frac{d\Phi}{dr} \right) \right|_{r=R}W(R)=0
\end{equation}

The real part of the complex eigenfrequencies $\omega$ give the frequency of the perturbation functions.

\section{Results}
\label{sec:res}

\subsection{Testing the numerical scheme}
\label{sec:initest}

\begin{figure} [htbp]
\centering
\includegraphics[height=8 cm,width=10 cm]{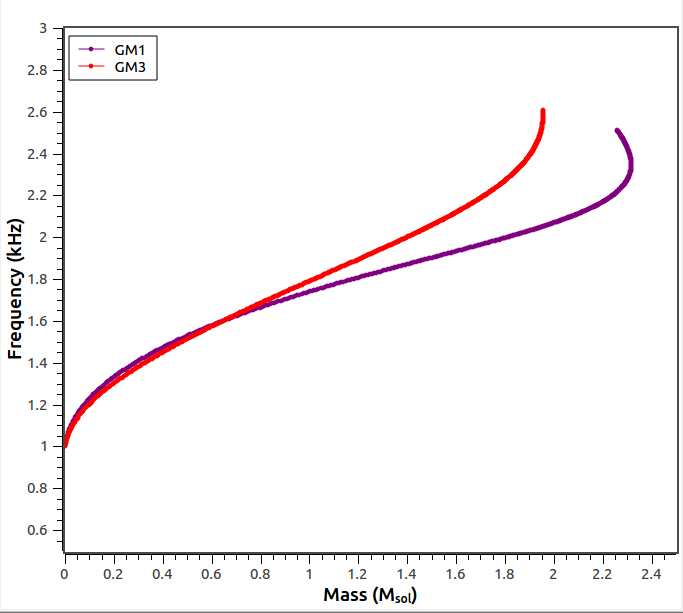}
\caption{$f$-mode frequencies for the parameter sets GM1 and GM3}
\label{fig:fmodes_gm}
\end{figure}

\indent In order to test the numerical scheme, we first reproduce well known results for $f$-mode frequencies for the GM1 parameter set (see e.g., \cite{Vasquez,Sandoval}) and also for the GM3 set. In Figure \ref{fig:fmodes_gm} we display the fundamental $f$-mode frequencies for these two reference parameter sets. As expected, the $f$-modes have frequencies within 1-3 kHz compatible with previous results reported in the literature.
\\

\subsection{Sensitivity study}
\label{sec:sens}

\subsubsection{Calculation of $l=2$ $f$-modes}
\label{sec:fmodes}

\indent  In Sec.~\ref{sec:formalism}, we discussed the uncertainties associated with the nuclear saturation parameters, which in turn result in the uncertainty in the EoS. Having tested the numerical scheme of the $f$-mode frequencies in Sec.~\ref{sec:initest}, we now extend it to investigate the entire parameter space. We vary each of the nuclear saturation parameters individually within their known uncertainties listed in Table 1 and study the sensitivity of the $f$-mode frequencies to each of the variations.

\begin{figure}[htbp]
  \begin{center}
     \includegraphics[width=.45\textwidth]{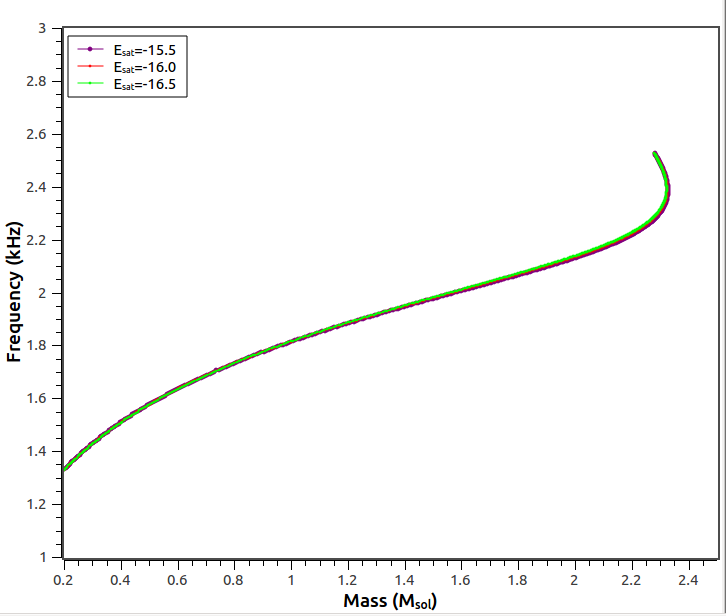}
      \includegraphics[width=.45\textwidth]{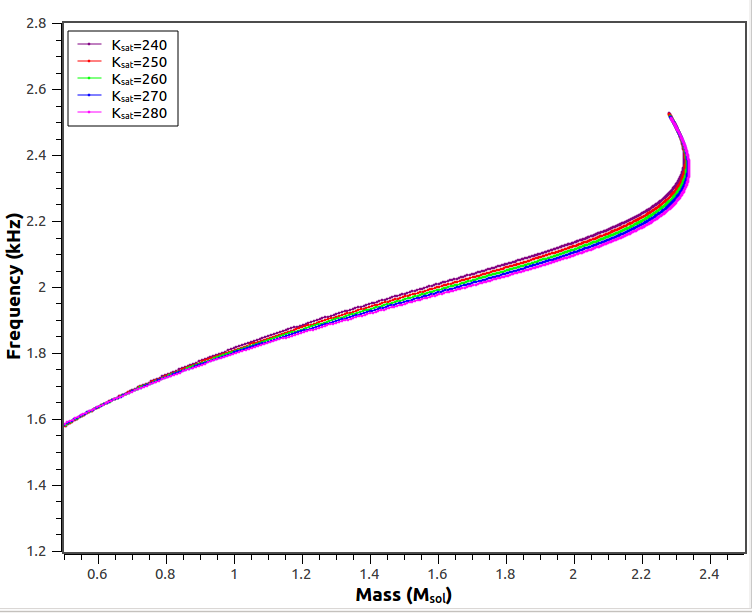}
      \includegraphics[width=.45\textwidth]{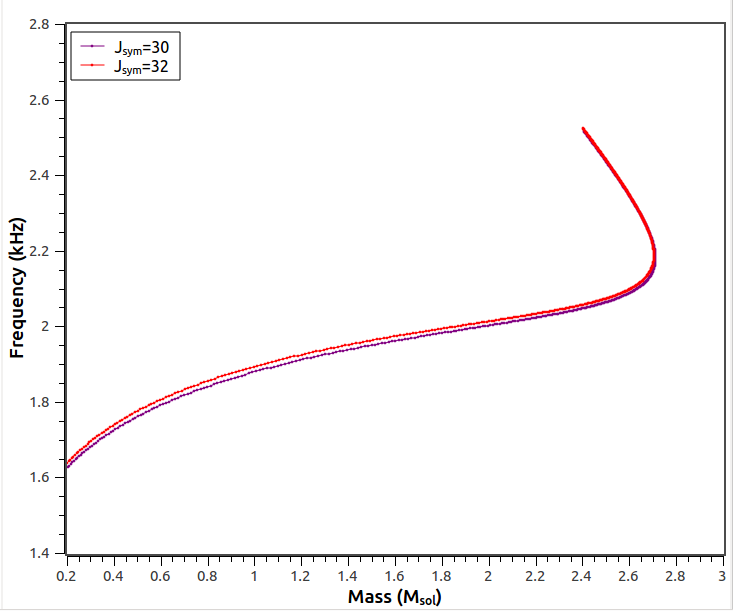}
      \includegraphics[width=.45\textwidth]{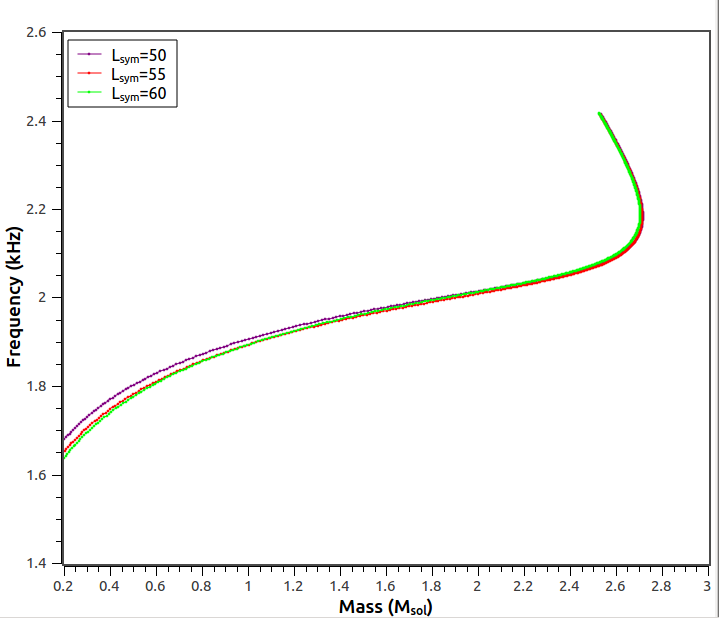}
    \caption{$f$-mode frequencies as a function of NS mass for EoSs with varying empirical parameters: energy at saturation $E_{sat}$ (upper left panel), compressibility $K_{sat}$ (upper right), symmetry energy $J_{sym}$ (bottom left) and slope of symmetry energy $L_{sym}$ (bottom right)}
    \label{fig:fmodes_vary}
  \end{center}
\end{figure}

\begin{figure} [htbp]
\centering
\includegraphics[height=8 cm,width=10 cm]{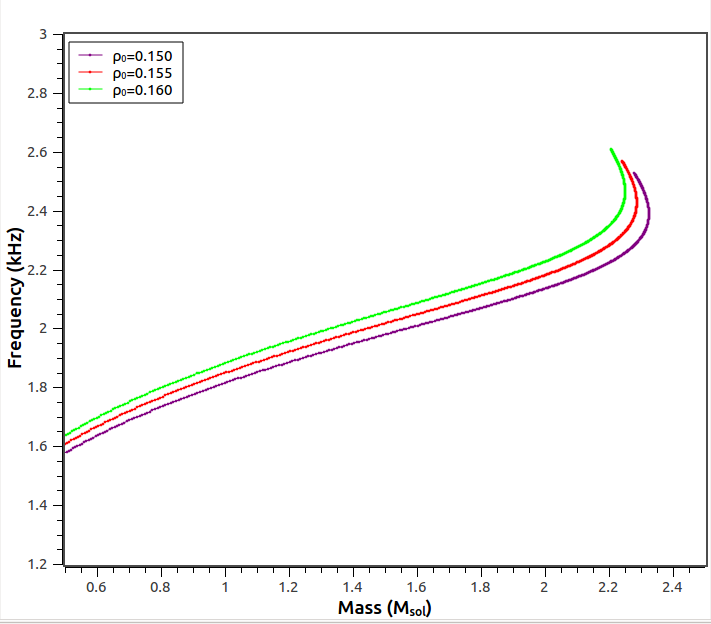}
\caption{$f$-mode frequencies as a function of mass for EoSs with varying saturation density $\rho_{sat}$}
\label{fig:fmodes_rhosat}
\end{figure}

\begin{figure} [htbp]
\centering
\includegraphics[height=8 cm,width=10 cm]{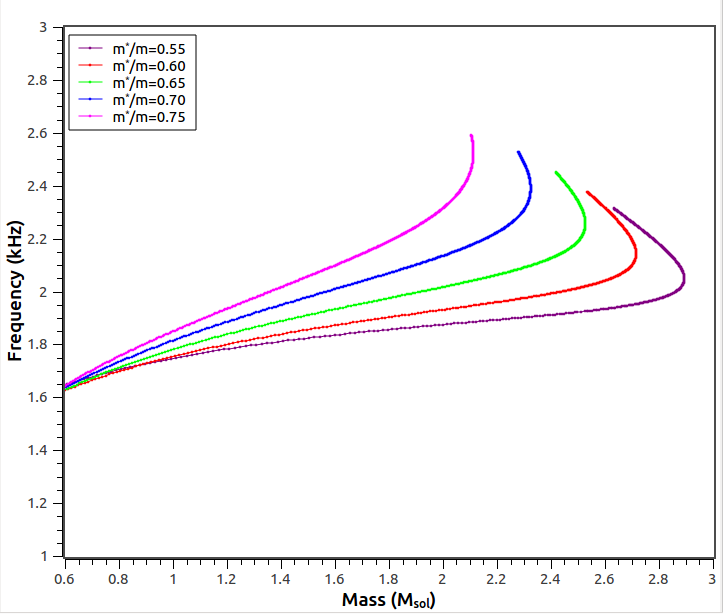}
\caption{$f$-Mode Frequencies for EoSs with varying effective masses $m^*/m$}
\label{fig:fmodes_effm}
\end{figure}

It is obvious from the panels in Fig.~\ref{fig:fmodes_vary} that the influence of varying the isoscalar parameters, energy per particle at saturation $E_{sat}$ and compressibility $K_{sat}$, is negligible. Similarly, the variation of the isovector parameters, symmetry energy $J_{sym}$ and its slope $L_{sym}$, do not vary the $f$-mode frequencies significantly, as can be seen from the same figure. In Fig.~\ref{fig:fmodes_rhosat}, we find a small non-zero effect of the variation in the saturation density $\rho_{sat}$. However, the most important influence on the frequencies comes from the variation in the effective nucleon mass $m^*/m$. In Fig.~\ref{fig:fmodes_effm}, we can clearly differentiate between the different EoSs with varying effective mass in the $f$-modes as a function of the stellar mass. The frequencies as a function of NS masses are seen to vary between 2 to 2.6 kHz, and the variation is monotonic with increasing $m^*/m$. This could have interesting consequences for extracting dense matter physics from the detection of $f$-mode frequencies for known stellar masses.
\\

\subsubsection{$f$-modes and tidal deformability}
\label{sec:tidal}

Among the various NS astrophysical observables that can help to constrain the nuclear EoS, one of the most promising quantities that has recently emerged is the tidal deformability. With the discovery of gravitational waves from mergers of NSs, it was seen that during the inspiral phase the NSs exert strong gravitational forces on each other, and the deformation produced depends on their EoS \cite{Abbott2017a}. The relation between the dimensionless tidal deformability and the NS compactness $C=M/R$ is given by
$\Lambda = \frac{2k_2}{3C^5}$, where $k_2$ is the second tidal Love number.
\\

\begin{figure} [htbp]
\centering
\includegraphics[height=8 cm, width=10cm] {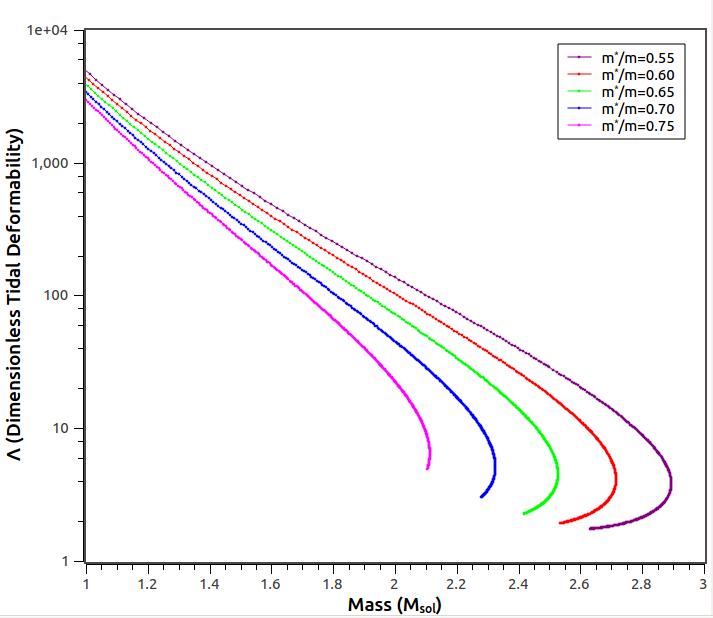}
\caption{Dimensionless tidal deformability $\Lambda$ as a function of NS mass (in $M_{sol}$) for different effective nucleon masses $m^*/m$ }
\label{fig:tidalm}
\end{figure}

\begin{figure} [htbp]
\centering
\includegraphics[height=8 cm, width=10cm] {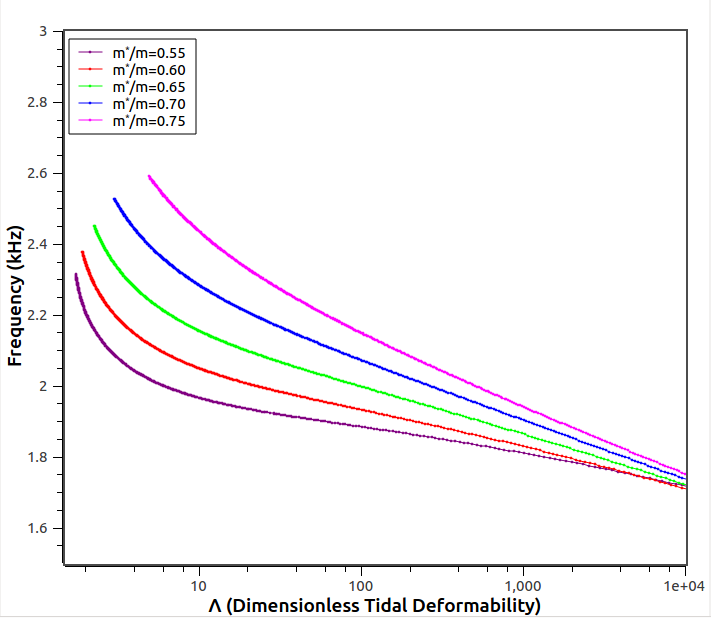}
\caption{$f$-mode frequencies as a function of dimensionless tidal deformability $\Lambda$ for varying $m^*/m$}
\label{fig:freqtidal}
\end{figure}

In order to estimate the influence of the $m^*/m$, the dominant empirical parameter that affects the $f$-mode frequencies, we first demonstrate its effect on the EoS and tidal deformability. In Fig.~\ref{fig:tidalm}, we plot the dimensionless tidal deformability $\Lambda$ as a function of NS mass (in $M_{sol}$). We see that the variation of effective nucleon masses $m^*/m$ influences $\Lambda$ as a function of the NS mass, and the range of values is consistent with recent observations \cite{Abbott2020,De}.
\\

Now, we investigate the influence of variation in $m^*/m$ on the $f$-mode frequencies as a function of $\Lambda$ in Fig.~\ref{fig:freqtidal}. We find that the variation in effective nucleon masses causes a change in $f$-mode frequencies in the range (2-2.6 kHz), for a corresponding change in dimensionless tidal deformability in the range $\sim$ 1-10. From the curves, one can obtain the frequencies corresponding to the lower limit of tidal deformability ($\sim 160$, constraints from terrestrial nuclear experiments), and the upper limit of tidal deformability of NSs extracted from the GW170817 event by LIGO and VIRGO Collaborations ($\sim 580$) which can provide an interesting constraint for the effective nucleon mass and hence the nuclear EoS \cite{BaoAnLi}.
\\


\begin{figure} [htbp]
\centering
\includegraphics[height=8 cm, width=10cm] {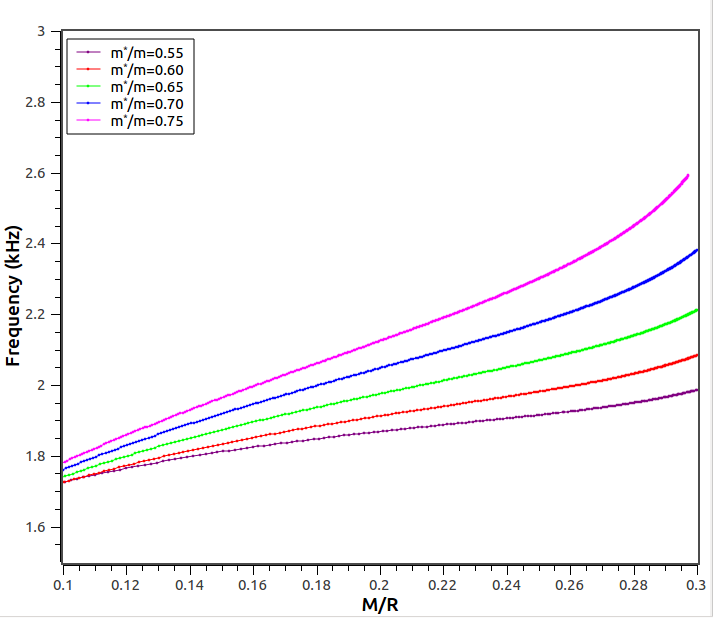}
\caption{$f$-mode frequencies as a function of stellar compactness $C=M/R$ for varying $m^*/m$}
\label{fig:freqcompact}
\end{figure}

\begin{figure} [htbp]
\centering
\includegraphics[height=8 cm, width=10cm] {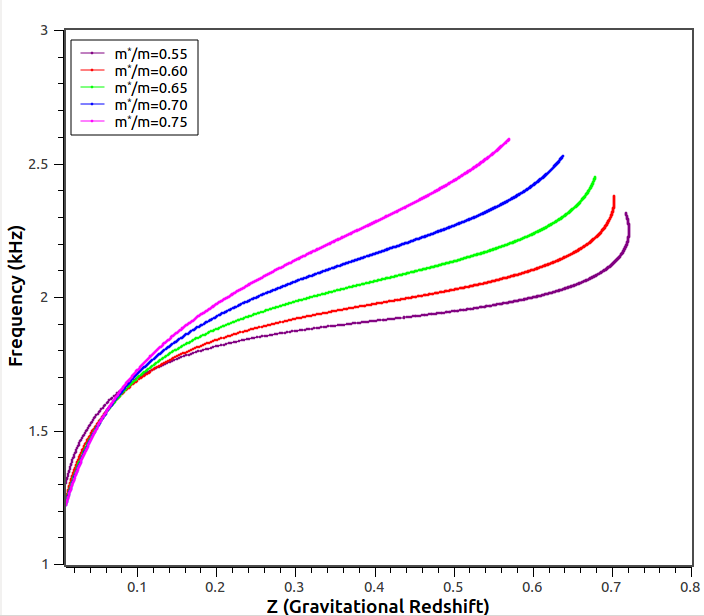}
\caption{$f$-mode frequencies as a function of gravitational redshift $Z$ for varying $m^*/m$}
\label{fig:freqZ}
\end{figure}

We already mentioned in the previous section that the tidal deformability is related to the stellar compactness. In order to use GW observations to estimate the NS mass and radius and to differentiate between different families of EoS, empirical relations between the frequency of $f$-modes and the compactness of the star may be useful \cite{Andersson96,Andersson98,Benhar}. We study in Fig.~\ref{fig:freqcompact} the effect of variation of $m^*/m$ on the $f$-mode frequencies as a function of $C=M/R$. Alternatively, one may also obtain the compactness from observations of the gravitational redshift $Z$ from spectral lines, as they are related by $Z=(1-2C)^{-1/2} - 1$. The corresponding variation of $f$-mode frequencies as a function of $Z$ are also shown in Fig.~\ref{fig:freqZ}.
\\

\subsection{Asteroseismology relations for $f$-modes}
\label{sec:astero}

\begin{figure} [htbp]
\centering
\includegraphics[height=8 cm, width=10cm] {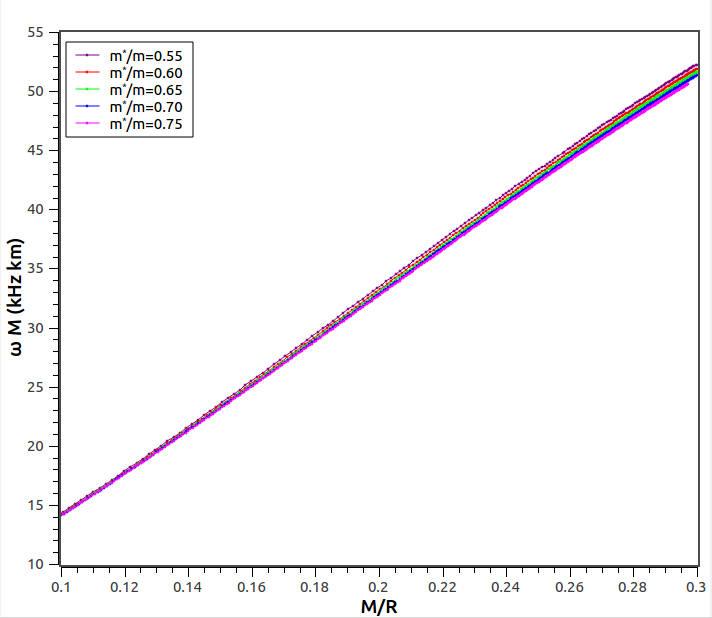}
\caption{Scaled $f$-mode frequency as function of stellar compactness $C$ for varying $m^*/m$}
\label{fig:wMcompact}
\end{figure}

\begin{figure} [htbp]
\centering
\includegraphics[height=8 cm, width=10cm] {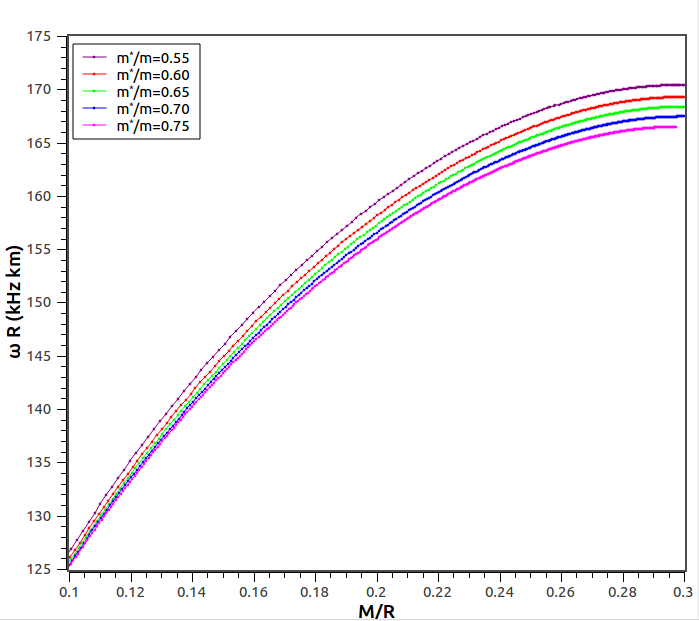}
\caption{Scaled $f$-mode frequency as function of stellar compactness $C$ for varying $m^*/m$}
\label{fig:wRcompact}
\end{figure}

In the previous Sec.~\ref{sec:tidal}, we studied the correlations between the $f$-mode frequency and stellar compactness $M/R$. Now we investigate some universal relations relating the scaled $f$-mode frequency versus compactness of NSs \cite{Andersson98}. Correlations of the $f$-mode frequency and damping time (or appropriately scaled by stellar mass and radius) with compactness and tidal deformability have been recently investigated \cite{BaoAnLi,Salcedo}. In Fig.~\ref{fig:wMcompact} we plot the frequency (scaled by the NS mass in km) and Fig.~\ref{fig:wRcompact} the frequency $\omega$ (scaled by the NS radius in km) as a function of compactness. We find that $\omega M$ as a function of $C$ retains a universal behaviour while $\omega R$ shows slight deviation from universal behaviour with variation in effective nucleon mass. 
\\

If we want to use GW data from NSs to estimate global properties like radius or mass, we would require empirical relations as much independent of the underlying EoS as possible. It was already pointed out in previous studies \cite{Schutz,Doneva2013,Benhar} that $f$-mode frequencies show a linear behaviour as a function of square-root of the mean density $(M/R^3)^{1/2}$. However such studies considered polytropic EoSs \cite{Andersson96} or realistic EoSs with only a few representative parameter sets \cite{Andersson98,Doneva2013}. Many of the chosen EoSs are not consistent with recent astrophysical data, such as the maximum mass constraint of 2$M_{sol}$. Linear fits derived from such studies would then introduce an error in obtaining relations useful for asteroseismology.
\\

\begin{figure} [htbp]
\centering
\includegraphics[height=8 cm,width=10 cm]{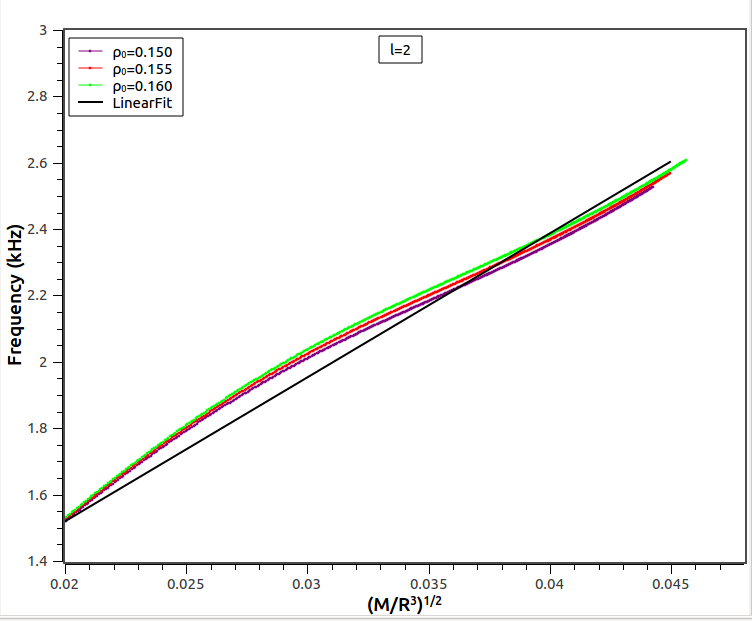}
\caption{$l=2$ $f$-mode frequencies as a function of $(M/R^3)^{1/2}$ for varying $\rho_{sat}$. The black solid line gives the linear fit to the curves.}
\label{fig:l2fit_rhosat}
\end{figure}

\begin{figure} [htbp]
\centering
\includegraphics[height=8 cm,width=10 cm]{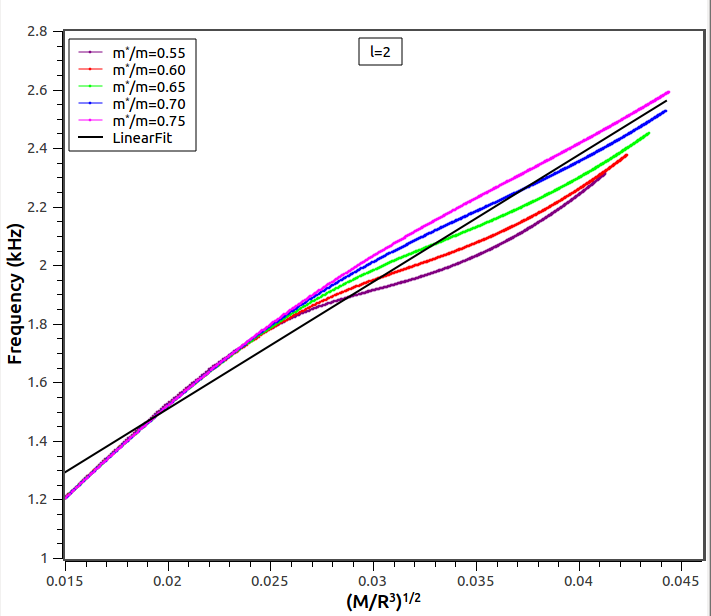}
\caption{$l=2$ $f$-mode frequencies as a function of $(M/R^3)^{1/2}$ for varying $m^*/m$}
\label{fig:l2fit_effm}
\end{figure}

In Sec.~\ref{sec:sens}, we already determined the key nuclear parameters that control the $f$-mode amplitudes. In Figs.~\ref{fig:l2fit_rhosat} and \ref{fig:l2fit_effm}, we now plot the $l=2$ $f$-mode frequencies as a function of $(M/R^3)^{1/2}$ for varying $\rho_{sat}$ and $m^*/m$. For each of the plots, we obtain a linear fit to the curves, marked by the bold black line.  The equations for the fits are as follows:\\
(i) For variation in saturation density:
$$y = 0.6499 + 1.6248 x $$
(ii) For variation in effective mass:
$$y = 0.6397 + 1.6243 x$$
where $y$ is the $l=2$ $f$-mode frequency (in kHz) as a function of $x = (\bar{M}/\bar{R}^3)^{1/2}$, in terms of the dimensionless variables $\bar{M} = M/(1.4 \> M_{sol})$ and $\bar{R} = R/(10 \> km)$ \cite{Doneva2013}. 
By inverting such relations, the mass and the radius of each NS can then be calculated upto the accuracy of the average density and the stellar compactness.
\\

\subsubsection{Higher order $f$-modes}
\label{sec:higherfmodes}

\begin{figure} [htbp]
\centering
\includegraphics[height=8 cm,width=10 cm]{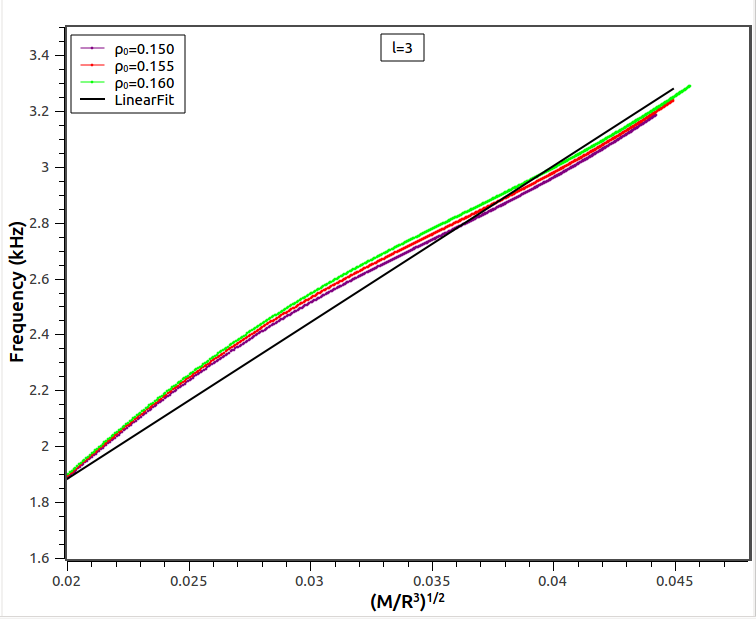}
\caption{Same as Fig.~\ref{fig:l2fit_rhosat} but for higher order $l=3$ mode}
\label{fig:l3fit_rhosat}
\end{figure}

\begin{figure} [htbp]
\centering
\includegraphics[height=8 cm,width=10 cm]{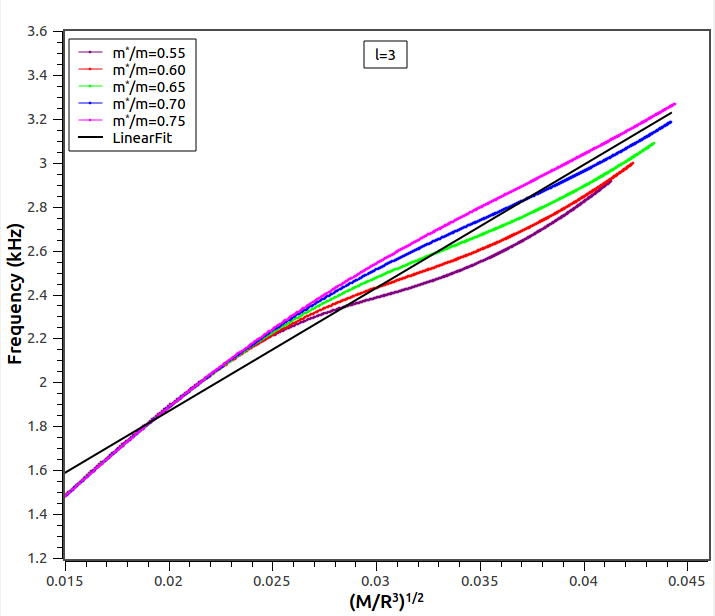}
\caption{Same as Fig.~\ref{fig:l2fit_effm} but for higher order $l=3$ mode}
\label{fig:l3fit_effm}
\end{figure}

\begin{figure} [htbp]
\centering
\includegraphics[height=8 cm,width=10 cm]{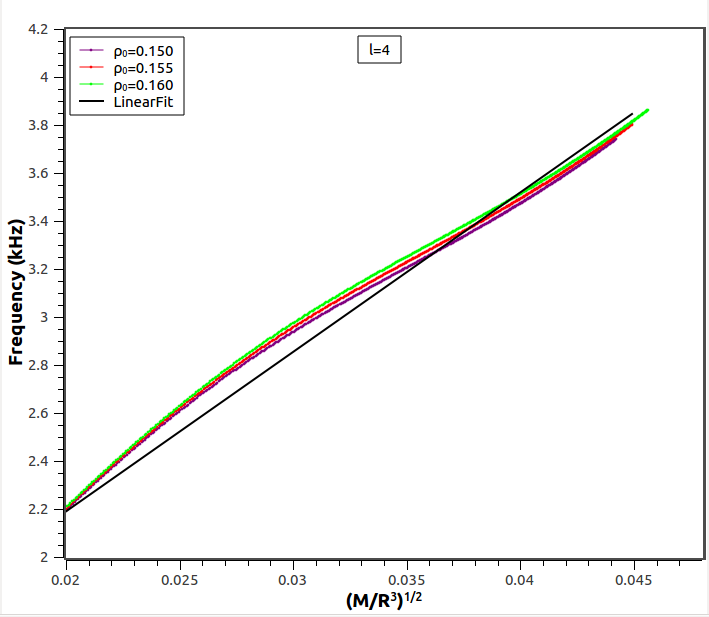}
\caption{Same as Fig.~\ref{fig:l2fit_rhosat} but for higher order $l=4$ mode}
\label{fig:l4fit_rhosat}
\end{figure}

\begin{figure} [htbp]
\centering
\includegraphics[height=8 cm,width=10 cm]{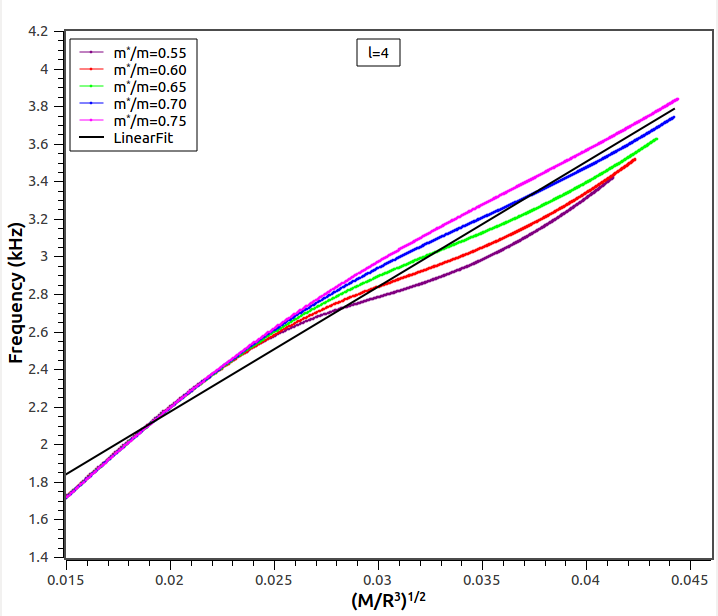}
\caption{Same as Fig.~\ref{fig:l2fit_effm} but for higher order $l=4$ mode}
\label{fig:l4fit_effm}
\end{figure}

 Recent calculations \cite{Gaertig2011,Passamonti2013} have investigated the instability window relevant to $f$-modes and concluded that higher order modes (e.g. $l=3,4$ modes) could be more dominant than the quadrupole $l=2$ mode. In analogy with $l=2$ case, we perform a study of the variation of the $f$-mode frequencies as a function of $(M/R^3)^{1/2}$ for the variation in $\rho_{sat}$ for higher orders in $l=3,4$. These are given in Figs.~\ref{fig:l3fit_rhosat} and \ref{fig:l4fit_rhosat}. 
The variation of $f$-mode frequencies with effective mass for higher orders are similarly plotted in Figs.~\ref{fig:l3fit_effm} and \ref{fig:l4fit_effm} respectively. We also obtained the linear fits to the curves for the $l=3$ case:\\
(i)  For variation in saturation density
$$y = 0.7628 + 2.9039 x$$
(ii) For variation in effective mass:
$$y = 0.7510 + 2.9032 x$$
where $y$ is the $l=3$ $f$-mode frequency (in kHz) as a function of $x = (\bar{M}/\bar{R}^3)^{1/2}$ defined previously.
\\

 Similarly, for the $l=4$ case, the linear fits are given by:\\
(i)  For variation in saturation density
$$y = 0.8586 + 2.4891 x$$
(ii) For variation in effective mass:
$$y = 0.8454 + 2.4882 x~.$$

\section{Discussions}
\label{sec:disc}

In this work, we studied the role of the underlying dense matter physics on $f$-modes. Within the framework of the RMF model, we performed a systematic investigation of the influence of the uncertainties in empirical nuclear EoS parameters, consistent with recent nuclear experimental data, on the $f$-mode frequencies. The equations of state obtained from these parameter sets were all compatible with the 2$M_{sol}$ maximum mass limit. We found the nucleon in-medium mass to be the most dominant parameter, while the saturation density has a small non-zero effect. Variation in effective mass between 0.55-0.75 resulted in a corresponding variation of $f$-mode frequencies between 2-2.6 kHz as a function of NS masses.
\\

We further investigated the effect of uncertainty in effective nucleon mass on the relation between $f$-mode frequencies and tidal deformability, compactness and gravitational redshift. These quantities can be derived and will be further constrained with the rapid improvement in multi-messenger astrophysical observations. We investigated whether the uncertainty in effective mass affects the universality in asteroseismology relations. We also obtained linear fits to correlations between $f$-mode frequencies and square root of the average density, for quadrupole ($l=2$) and higher order ($l=3,4$) modes.
\\

In order to extract information about $f$-modes in NS merger remnants, one must include rotation effects. We have not considered rotating NSs in this investigation, as it is beyond the scope of this undergraduate project. One must also consider the influence of exotic constituents of matter (hyperons, kaons, deconfined quark matter) on $f$-modes. Several works have attempted such an investigation, but again for a few selected EoSs \cite{Vasquez,Sandoval}. We are also currently performing such an investigation and the results will be reported in a forthcoming publication.
\\
 
With the recent detection of gravitational waves from binary compact objects, particularly mergers of binary neutron stars, prospects of constraining nuclear physics using gravitational waves have become attractive. It has been speculated that $f$-modes are among the most significant sources of GWs due to the Chandrasekhar-Friedman-Schutz (CFS) mechanism, for both isolated NSs or in the post-merger scenario. Recent studies suggest that GWs produced by unstable $l=m=2$ and the $l=m=4$ $f$-modes could be detectable by the future Einstein Telescope for sources in the Virgo cluster or for $l=m=3$ modes even by the LIGO/VIRGO \cite{Passamonti2013}. Then the possibility of distinguishing between NS EoSs using information about $f$-modes and NS global observables could prove be very interesting. 
\\

\section{Acknowledgements}
This short project was done entirely remotely during the recent Covid19 crisis. S. J. did a commendable job, being an undergraduate student, working on this paper under my supervision from home without any help. Bravo!\\
D. C. would like to thank the faculty, LIGO-India team and computing section at Inter-University Centre for Astronomy and Astrophysics, where this work was performed, for their full support and assistance during the crisis.
\\


\end{document}